\documentstyle[12pt,graphicx]{article}

\def\be{\begin{equation}}

\def\ee{\end{equation}}

\def\bea{\begin{eqnarray}}

\def\eea{\end{eqnarray}}

\begin{document}

\title{SPIN EXCITATIONS IN LOCALIZED AND ITINERANT MAGNETS}

\author{ B. ROESSLI\\ 
Laboratory for Neutron Scattering \\ Paul Scherrer Institut and ETH Zurich\\ CH-5232 Villigen, PSI,
Switzerland \\
\\
 P. B\"ONI\\
Physik-Department E21, Technische Universit\"at M\"unchen
\\ D-85748 Garching, Germany}




\maketitle
\abstract{Collective excitations in magnetic materials
can be investigated by means of inelastic neutron scattering. We
show that this experimental method gives access to the complete
spectrum of magnetic fluctuations through the energy- and
momentum-dependence of the dynamical susceptibility. We focus on
the dynamical properties of magnets with localized spin densities
and of metals. From such studies, microscopic parameters like
exchange integrals, spin-wave stiffness and relaxation times can
be determined. This is of great help to test current theories in
magnetism.}

\newpage

\section{Introduction}

Many materials exhibit a spontaneous phase transition from a
paramagnetic state to an ordered configuration of their magnetic
moments below some critical temperature. This cooperative
phenomenon is found in a wide range of materials, like insulators,
metals, superconductors, heavy fermion systems, and so on. Ordered
magnetic states can be classified as weak or strong ferromagnets,
antiferromagnets, ferrimagnets, and incommensurate structures.
Magnetic ordering originates from electrostatic interactions which
induce correlations between the electrons. An exact mathematical
treatment of the Coulomb interaction involving many electrons
together with the Pauli-principle has been, up to now, an
impossible task. Accordingly,  some simplifications of the problem
have to be made. An effective many-body Hamiltonian which takes
into account both the kinetic energy of the electrons and the
Coulomb interaction has been derived by Hubbard \cite{hubbard}.
Considering a single-electron band and neglecting inter-atomic
interactions, the Hubbard model is given by
\begin{equation}
H=-t\sum_{i,j}{}\sum_{\sigma}{(c^\dagger_{i\sigma}c^{\vphantom
\dagger}_{j\sigma} +c^{\vphantom
\dagger}_{i\sigma}c^\dagger_{j\sigma})}+
 U\sum_{j}n_{j\uparrow}n_{j\downarrow}.
\label{hubbard}
\end{equation}
The first term describes electrons hopping through the lattice,
while the interaction term describes the Coulomb repulsion between
electrons in the same orbital,
\begin{equation}
U=\int{d\vec r_1\int{d\vec r_2|\phi(\vec r_1)|^2{{e^2}\over{|\vec
r_1-\vec r_2|}} |\phi(\vec r_2 )|^2 }}.
\end{equation}
$|\phi(\vec r_{1,2})|^2$ are the charge densities.

If the kinetic energy dominates ($t>>U$), the Hubbard model
describes ferromagnetism in metallic systems. On the other hand,
in  the limit of large Coulomb interactions ($U>>t$) and at
half-filling Eq. \ref{hubbard} reduces to
\begin{equation}
H=J\sum_{i,j}\vec S_i\cdot \vec S_j,
\label{heisenberg}
\end{equation}
with $J$ equal to $J=4t^2/U$ (see Ref. \cite{fazekas}). Equation
\ref{heisenberg} is the Heisenberg model. It shows that the
magnetic interactions in solids can be described by the mutual
interaction of pairs of spins $\vec S$ which are coupled together
by the exchange integral $J$. Derived from the Hubbard model, Eq.
\ref{heisenberg} favors an antiferromagnetic ground state as the
energy is minimized for anti-parallel spins. However it has the
same form as the Hamilton operator originally derived by
Heisenberg for ferromagnets. The Heisenberg model is therefore
expected to be valid for positive and negative values of the
exchange integral $J$.\\ Neutron scattering is the only method
that allows to determine both the spatial and time correlations of
the magnetic excitations through the dipolar coupling of the
magnetic moment of the neutron with the unpaired electrons of the
sample,
\begin{equation}
{{d^2\sigma}\over{d\Omega dE}}= {k_f\over k_i} (\gamma r_0 {1\over
2}gF(\vec Q))^2 \sum_{\alpha \beta}{(\delta_{\alpha \beta}-{\hat
Q_\alpha}{\hat Q_\beta}) S^{\alpha \beta}(\vec Q, \omega)}.
\label{crose}
\end{equation}
$\gamma=-1.913$ is the gyromagnetic ratio, $r_0 =0.2818 \cdot
10^{-12}$ cm is the classical radius of the electron, g the
Land\'e splitting factor, $k_i$ and $k_f$ are the wavenumbers of
the incident and scattered neutrons, and $F(\vec Q)$ is the form
factor. $\hat Q$ is the scattering vector normalized to unity and
$S^{\alpha \beta}(\vec Q, \omega)$ the magnetic scattering
function. $S^{\alpha \beta}(\vec Q, \omega)$ is related to the
imaginary part of the wavevector and frequency dependent
susceptibility through the fluctuation-dissipation theorem by
\begin{equation}
S^{\alpha \beta}(\vec Q, \omega)={\hbar\over \pi}{1 \over
{(1-\exp{(-\hbar \omega/k_B T)})}} \Im\chi(\vec Q,\omega).
\end{equation}
This implies that the magnetic moment of the neutron acts as a
wavevector and frequency dependent magnetic field that probes the
dynamic magnetic response of the sample.

\section{Collective spin excitations in localized magnets}

\subsection{Spin Waves in the Ferromagnetic Heisenberg Model}
At zero temperature, every spin in a ferromagnet points along the
same direction, i.e. along the magnetization $\vec M$. The
ground-state can therefore be written as $|\rm
\uparrow\uparrow\uparrow...>$. The saturation magnetization of the
sample is accordingly given by $M_0=Ng\mu_BS$. Upon increasing the
temperature, the magnetization decreases indicating deviations of
the spins from their parallel alignment, i.e. $\rm S^z_j
\rightarrow S^z_j-1$. In the quantum-mechanical formalism, this is
achieved by the annihilation operator $\rm S^-_j$ that reduces the
spin component $\rm S^z$  at the spin position $j$. If $S=1/2$,
this yields $\rm S^-_{j} |\rm \uparrow\uparrow\uparrow...>= |\rm
\uparrow\uparrow\uparrow... \uparrow\downarrow\uparrow...>$. As
the spins are coupled by the exchange interaction, the excitation
will propagate through the crystal like a wave. Because the
crystal lattice is periodic in space, it is of advantage to
introduce the Fourier transform of the exchange interaction and
spin operator, namely $J(\vec r_{i,j})=J(\vec r_i - \vec
r_j)={1\over N}\sum_{\vec k} {J(\vec k) \exp{(i\vec k\cdot \vec
r_{i,j})}}, S_j^z=\sum_{\vec k}{S_{\vec k}^z \exp{(-\vec k \cdot
\vec r_{i,j})}}$, and $S^{+,-}_j=S_j^{x}\pm S_j^{y}=\sum _{\vec
k}{S_{\vec k}^{+,-}\exp{(\mp i\vec k \cdot \vec r_j)}}$.
Accordingly, the Heisenberg operator transforms into
\begin{equation}
H=N\sum_{\vec k}{J(\vec k)[S^z_{\vec k}S^z_{-\vec k}+ {1\over
2}(S^+_{\vec k}S^-_{\vec k}+S^-_{\vec k}S^+_{\vec k}})].
\end{equation}

Using the commutation relations between the spins
$[S^+_k,S^-_q]=2S^z_k\delta_{k,q}$ and $[S^{+,-}_k,S^z_q]=\mp
2S^{+,-}_k\delta_{k,q}$, the dispersion relation of the magnetic
excitations is given by the solution of the equation of motion
$i\hbar {\partial\over{\partial t}} S^-_k(t)=[S^-_k,H]$. This is
achieved within the 'linear' approximation by setting
$[S^+_k,S^-_q]\simeq \delta_{k,q}2S$, i.e. it is assumed that the
spin deviations from their maximal value are small. This approach
is valid at low temperatures $T\rightarrow 0$. As we will see
below, the 'linear' approximation breaks down when the magnetic
fluctuations are large, and in particular when the temperature is
close to the Curie temperature $T_c$. Making use of the
'\textit{Ansatz}' $S^-_{\vec k}(t)=e^{i\omega(\vec k) t}S^-_{\vec
k}$, the dispersion relation for the spin waves in an isotropic
ferromagnet is obtained
\begin{equation}
\hbar\omega(\vec k)=2S(J(0)-J(\vec k)). \label{ferrodisp}
\end{equation}
Equation \ref{ferrodisp} shows that the dispersion of the magnetic
excitations is a function of the Fourier transform of the exchange
integrals between the spins. For small values of $\vec k$, the
dispersion is quadratic $\hbar \omega (\vec k)=Dk^2$ with a
stiffness constant $D=2SJa^2$ ($a$ is the lattice constant). For a
cubic crystal $D$ has the same value for all directions of $k$.

\subsubsection{Neutron-Scattering Cross-Section}

The inelastic scattering of neutrons allows to measure the
dispersion relation $\omega(\vec k)$ given in Eq. \ref{ferrodisp}
directly, as  \cite{lovesey} the magnetic neutron cross section is
equal to (see Eq.~\ref{crose})
\begin{eqnarray}
{{d^2\sigma}\over{d\Omega dE}} & = & 0.2916 \cdot
10^{-24}{k_f\over k_i} ({1\over 2}gF(\vec Q))^2 \exp{(-2W(\vec
Q))}\nonumber \\ & &\times \{(1-Q^2_z){1\over
2\pi\hbar}\int_{-\infty}^{\infty}dt \exp(-i\omega t) \langle
S^z_{\vec Q}S^z_{-\vec Q}(t)\rangle \\
 &  & +{1\over 4}{(1+Q^2_z){1\over 2\pi\hbar}\int_{-\infty}^{\infty}dt \exp(-i\omega t)
 \langle S^+_{\vec Q}S^-_{\vec Q}(t)}+S^+_{-\vec Q}S^-_{-\vec Q}(t)\rangle\} \nonumber.
\label{cross-section}
\end{eqnarray}
Here, $W(\vec Q)$ is the Debye-Waller factor. The scattering
vector $\vec Q$ is related to the spin-wave wave-vector $\vec k$
through the relation $\vec Q=\vec \tau + \vec k$, where $\vec
\tau$ is a reciprocal lattice vector. The neutron-scattering
cross-section for the isotropic ferromagnet is obtained by
evaluating the thermal averages $\langle \dots \rangle$ in Eq. 8.
Within the spin-wave approximation, $\langle
S^z_{\vec Q}S^z_{-\vec Q}(t)\rangle$ does not depend on time and
hence gives no contribution to the inelastic scattering, while
$\langle S^+_{\vec Q}S^-_{\vec Q}(t)\rangle = 2SN[\exp({{\hbar
\omega} \over {k_BT}})-1]^{-1}=2SNn_{\vec k}$ \cite{lovesey}.
Accordingly, Eq.  8 
 reduces to
\begin{eqnarray}
{{d^2\sigma}\over{d\Omega dE}} & = & 0.2916 \cdot
10^{-24}{k_f\over k_i} ({1\over 2}gF(\vec Q))^2 \exp(-2W(\vec Q))
\nonumber \\
 &  & \times {1\over 2}S {{(2\pi)^3}\over {v_0}} \sum_{\vec k, \vec \tau, \pm}^{}(n_{\vec k}+{1\over 2}\pm {1\over 2})
\delta(\hbar \omega_{\vec k} \mp \hbar \omega) \delta(\vec Q \mp
\vec k - \vec \tau).
\label{ferrocs}
\end{eqnarray}

\subsubsection{Example}

\begin{figure}

\begin{center}

\includegraphics*[scale=0.25]{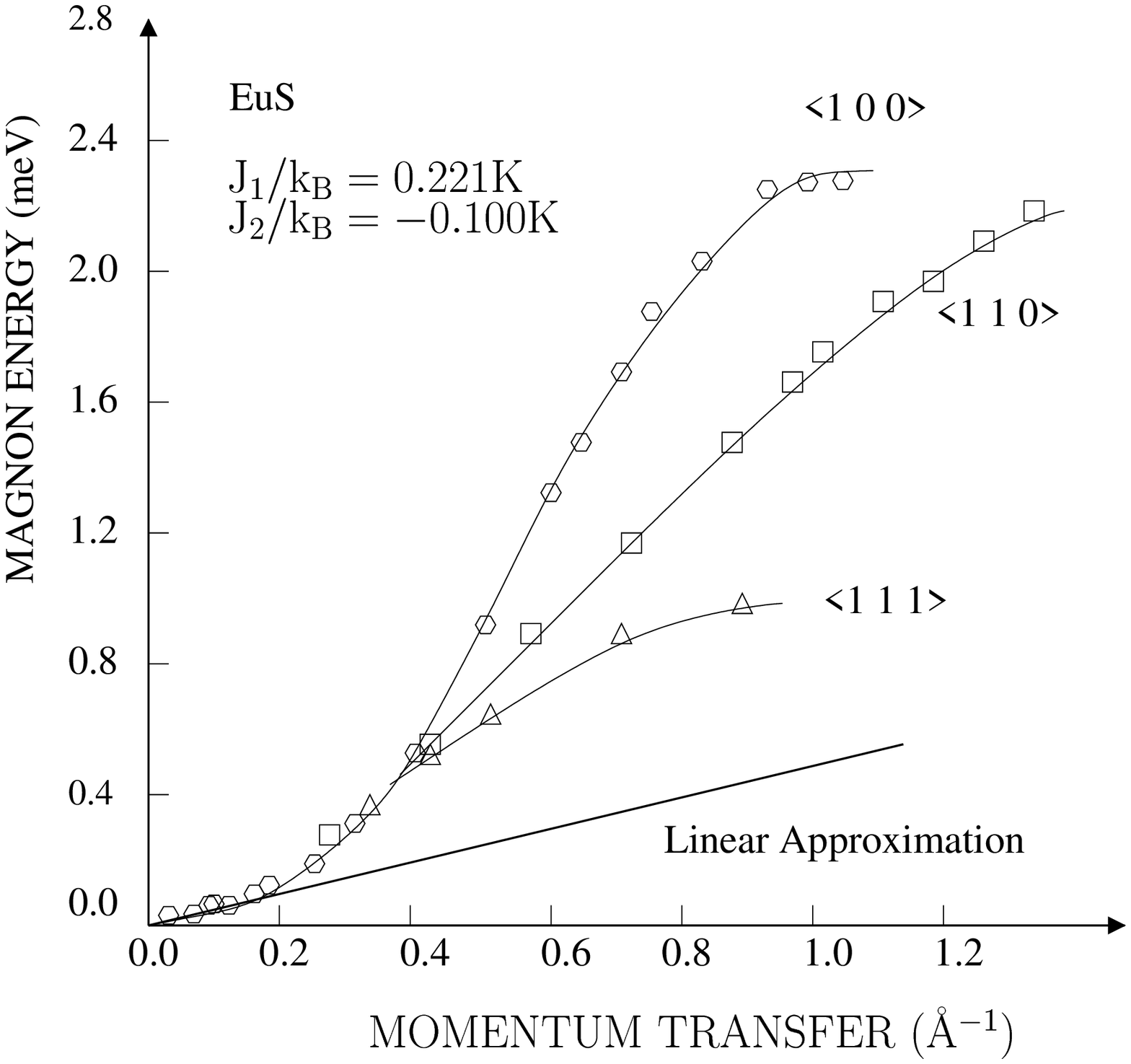}
\kern 0.5cm
\includegraphics*[scale=0.5]{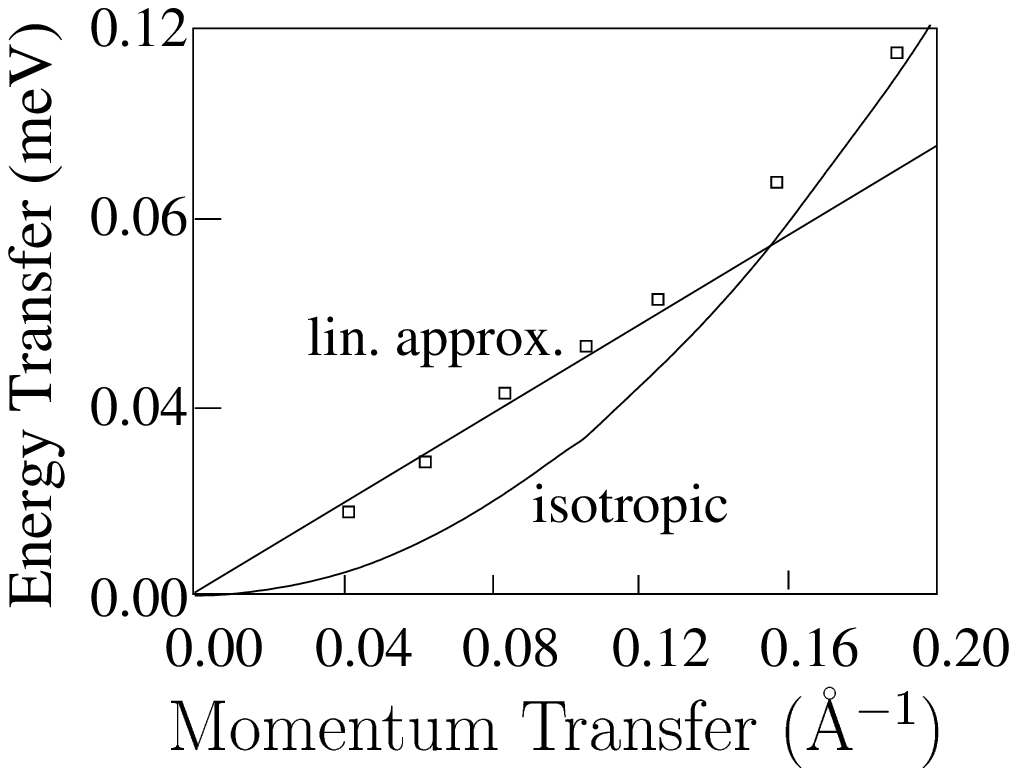}
\vskip 4pt \caption{\textit{Left}: Dispersion of the spin waves in
EuS. \textit{Right}: Expanded scale of the magnon dispersion in
EuS, demonstrating the linear dispersion at small $k$ values
(after Refs. \cite{Bohn,boni}).}
\label{EuS}
\end{center}
\end{figure}

The insulators EuS and EuO are typical examples of ferromagnets
which are well described by the isotropic Heisenberg Hamiltonian.
The electronic configuration of the 4\textit{f} electrons of $\rm
Eu^{2+}$ has a $^8S_{7/2}$ ground state and hence the magnetic
moment is only due to the spin. These two insulators undergo a
phase transition to an ordered ferromagnetic ground-state at $\rm
T_c$=16.6K and $\rm T_c$=69K, respectively. Results of spin-wave
measurements and the corresponding spin-wave dispersion in EuS
using the triple-axis method are shown in Fig.\ref{EuS} (taken
from Refs. \cite{Bohn,boni}).

The Fourier transform of the exchange interactions
\begin{equation}
 J(\vec q)=\sum_{i,j}J_{i,j}\exp (i \vec q \cdot \vec r_{i,j})
\end{equation}
 can be
easily calculated for a fcc lattice. The results of the
calculations using Eq. \ref{ferrodisp} are given by the solid line
in Fig. \ref{EuS}. As expected, nearest-neighbor exchange
interactions are strongest in EuS. However, magnetic interactions
of a spin at an $\rm Eu^{2+}$ position with more distant spins is
significant. Also, we point out that the sign of the magnetic
interaction changes as a function of distance which suggests
competing ferro- and antiferromagnetic exchange interactions in
EuS.\\ The value of the $\rm Eu^{2+}$ magnetic moment at
saturation is large, $\mu \sim 7 \mu_B$, and dipolar interactions
cannot be neglected. As dipolar interactions decay like $1/r^3$
and are anisotropic, they modify the spectrum of the spin-wave
excitations~\cite{holstein}:
\begin{equation}
E(\vec k) =\{\hbar \omega(\vec k)[\hbar \omega(\vec k) +
g\mu_B\mu_0 M(T) \sin^2\theta_{\vec k}]\}^{1/2}, \label{ferrodip}
\end{equation}
$\hbar \omega(\vec k)$ denotes the dispersion relation of Eq.
\ref{ferrodisp}; $\mu_0$ is the induction constant; $M(T)$ the
magnetization and $\theta_{\vec k}$ the angle between the
magnetization vector $\vec M$ and $\vec k$. At small $\vec k$ the
dipolar interactions lead to a linear dispersion
\begin{equation}
E(\vec k \rightarrow 0)\simeq k\sqrt{Dg\mu_B\mu_0M}\cdot
\sin\theta_{\vec k}.
\end{equation}
Fig. \ref{EuS} shows results of a high-resolution experiment that
demonstrates the linear dispersion in EuS due to dipolar
interactions \cite{boni}.

\subsubsection{Longitudinal fluctuations}
\label{sectionlong}

\begin{figure}
\begin{center}
\includegraphics*[scale=0.7,angle=-90.]{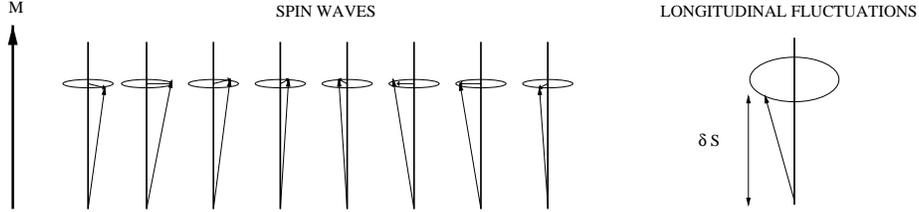}
\caption{Schematic representation of a spin-wave excitation
transverse to the magnetization $\vec M$
 (\textit{left}) and of longitudinal fluctuations (\textit{right}).}
\label{swandlong}
\end{center}
\end{figure}
Spin-wave theory is based on the assumption that spin deviations
from the equilibrium value are small and that the 'length' of the
spin vector $\vec S$ is a constant. This leads to the view that
spin waves are unperturbed collective modes \textit{transverse} to
the magnetization $\vec M$. Close to the phase transition and in
the paramagnetic phase, interactions between the spin-wave modes
are important and fluctuations along the magnetization vector
appear, as shown in Fig. \ref{swandlong}. In the temperature
regime below $T_c$ where critical fluctuations are large,
mode-mode coupling theory \cite{schwabl} predicts that the
transverse spin components still disperse like $\omega(\vec
k)=Dk^2$. The stiffness constant is given by $D=(4\pi
\xi/a)^{-1/2} \sqrt{TJ}/\hbar a^2$ ($\xi$ is the correlation
length and $a$ the lattice constant). In addition, the spin waves
are damped in such a way that the $\delta$-function in the neutron
cross-section of Eq. \ref{ferrocs} becomes a Lorentzian,
\begin{equation}
S(\vec k, \omega)=\omega(1-\exp{(-\omega/k_BT}))^{-1} \chi^T_{\vec
k}{\Gamma (\vec k) \over{(\omega\pm\omega(\vec k))^2+\Gamma (\vec
k)^2}}
\end{equation}
where $\chi^T_{\vec k}=4\pi M^2\xi N/(k^2k_BTa^3)$ is the
transverse susceptibility and $\Gamma=Ak^{2.5}\xi^{3/2}$. For
small wave-vectors $k$, the longitudinal part of the dynamical
susceptibility is a diffuse central peak of Lorentzian-shape
centered around zero-energy transfer and of width $\Gamma(\vec
k)\sim k^{2.5}$,
\begin{equation}
S(\vec k, \omega)=\omega(1-\exp{(-\omega/k_BT}))^{-1} \chi^L_{\vec
k}{\Gamma(\vec k)\over{\omega^2+\Gamma(\vec k)^2}}.
\end{equation}
The static longitudinal susceptibility $\chi^L(\vec k)$ behaves
like $c_1{{\xi/\gamma}\over{k}} +(1-c_1)(\xi/\gamma)^2$, where
$\gamma \sim 1$ \cite{schwabl}. Measuring the longitudinal
fluctuations with neutrons is however a difficult task, as when
the temperature gets close to $T_c$, the spin-waves renormalise
and merge with the central component. The best way to separate the
longitudinal from the transverse components is to use polarization
analysis, as explained in the next chapter.

\subsubsection{Polarization analysis}

\begin{figure}
\begin{center}
\includegraphics*[scale=0.4]{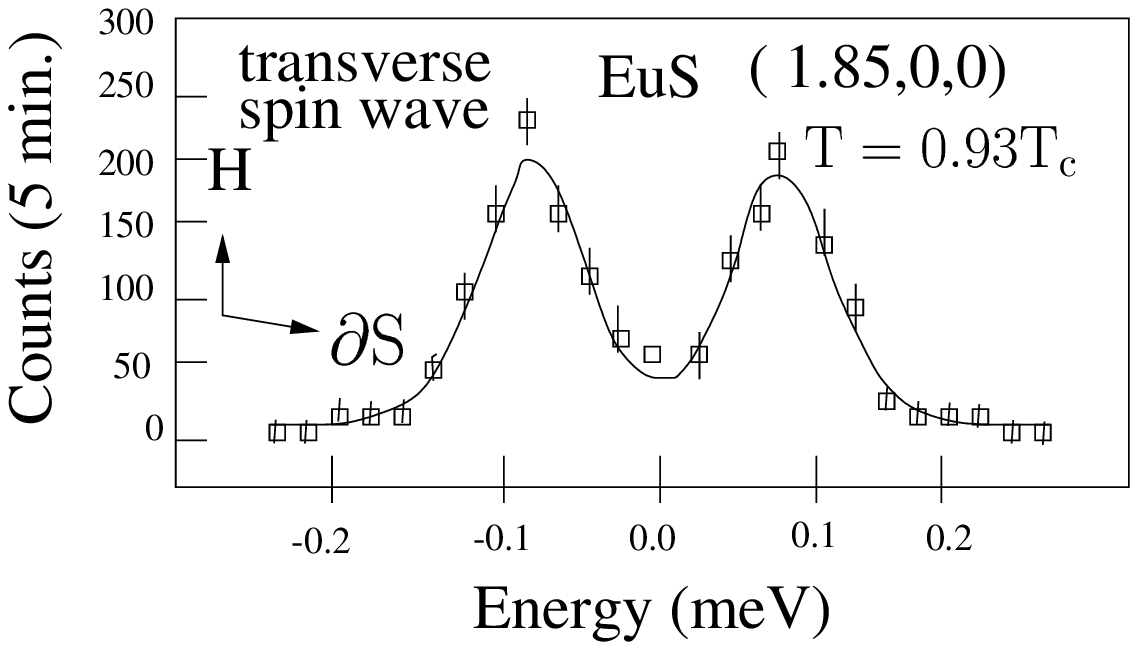}
\kern 0.5cm
\includegraphics*[scale=0.4]{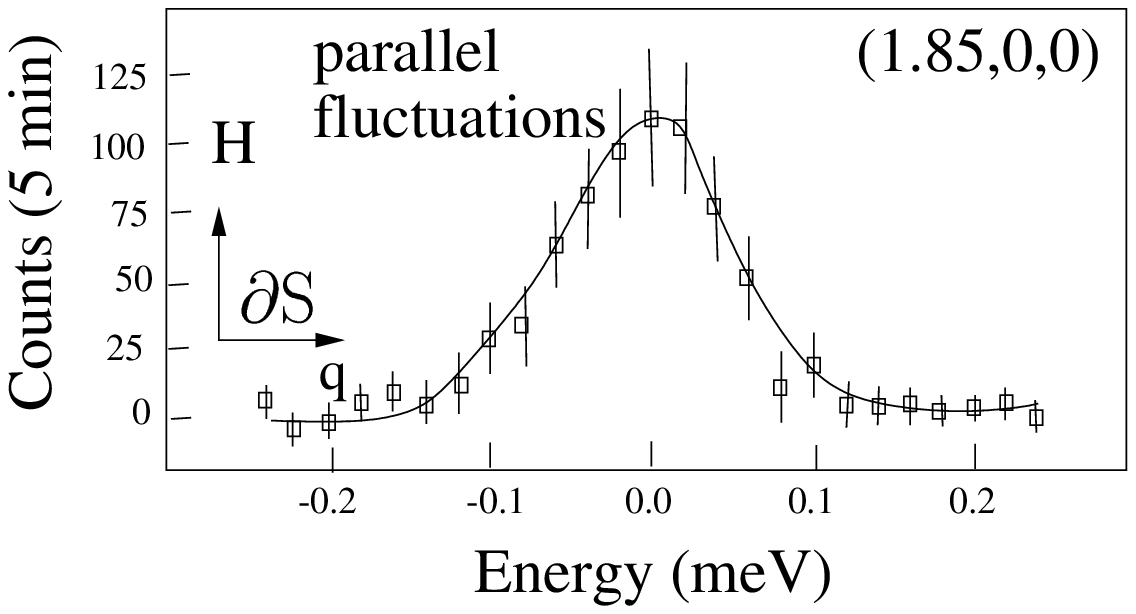}
\vskip 4pt \caption{\textit{Left}: Spin waves in EuS, spin-flip
channel. \textit{Right}: Longitudinal fluctuations in EuS, non
spin-flip channel (Taken from Ref. \cite{boni2}).} \label{sf}
\end{center}
\end{figure}

In the following section, we shall give a brief description of how
polarization analysis can be used to separate the transverse and
longitudinal components of the dynamical susceptibility in
ferromagnets. A more detailed discussion of the spin-dependent
neutron cross-sections can be found in the book of
Lovesey~\cite{lovesey} and in the seminal paper of Blume
\cite{blume}. The neutron cross-section for magnetic scattering
contains terms that depend on the spin of the neutron. The term
\textit{polarization analysis} refers to the analysis of the
spin-state of the neutron before and after the scattering process.
This implies that for most practical cases (but not all!), the
spin of the neutron before scattering has to be in a well defined
state. The neutron cross-section for scattering by spin waves in a
ferromagnet with a polarized beam is given by
\begin{equation}
{{d^2\sigma}\over{d\Omega dE}} \propto [1+(\hat{ \vec Q} \cdot
\hat{ \vec m})^2 \pm 2(\vec P_0 \cdot \hat{ \vec Q})(\hat{ \vec m}
\cdot \hat{ \vec Q})]
\end{equation}
where $\hat {\vec m}$ is a unit vector directed along the
magnetization and $\vec P_0$ is the polarization vector of the
neutrons before scattering by the sample. The $\pm$ sign refers to
spin-wave creation and annihilation, respectively. The final
polarization of the neutron beam $\vec P$, i.e. after scattering,
is \cite{moon}
\begin{equation}
\vec P={{\mp 2 \hat {\vec Q}(\hat {\vec Q} \cdot \hat {\vec
m})-\vec P_0[1+(\hat {\vec Q} \cdot \hat {\vec m})^2]+2\vec
M_x(\vec M_x \cdot \vec P_0)+2 \vec M_y(\vec M_y \cdot \vec
P_0)}\over {1+(\hat {\vec Q }\cdot \hat{\vec m})^2 \pm 2(\vec P_0
\cdot \hat{\vec Q})(\hat{\vec Q} \cdot \hat{\vec m}) }}
\label{polar}
\end{equation}
with $\vec M_x=\hat{\vec x} -(\hat {\vec x} \cdot \hat{ \vec
Q})\hat {\vec Q}$ and a similar definition can be written for
$\vec M_y$. $\hat{ \vec x}$ and $\hat {\vec y}$ are chosen to be
perpendicular to $\hat {\vec m}$. Equation \ref{polar} tells us
that the final polarization of the neutron depends on the
direction of its initial polarization and on the relative
orientation of the magnetization with the scattering vector $\vec
Q$. In particular, if an external magnetic field is applied
perpendicular to the scattering plane and is strong enough to
align the magnetic domains, both the magnetization vector and the
initial polarization are aligned along the field. In that case,
the spin of the neutrons will be reversed when neutrons are
scattered by the spin waves. On the contrary, the longitudinal
fluctuations will not influence the spin of the neutrons.
Therefore, for that particular scattering geometry, the spin waves
in EuS will occur in the \textit{'spin-flip channel'}, while the
longitudinal fluctuations will be \textit{'non spin-flip'}, as
shown in Fig. \ref{sf}.

\subsection{Ferrimagnets and antiferromagnets}

So far we have considered magnetic systems which contain one ion
in the magnetic cell. For compounds with many magnetic
sublattices, like ferrimagnets and antiferromagnets, the spectrum
of the magnetic excitations has more than one dispersion branch
$\omega(\vec k)$, like phonons in crystals. As an example, we will
consider here the spin-wave spectrum of $\rm CuFe_2O_4$. Cubic
$\rm CuFe_2O_4$ crystallizes in the spinel structure (space group
$Fd\bar{3}m$) with eight sites of tetrahedral symmetry (A-sites)
and sixteen sites of octahedral symmetry in the unit cell. The
magnetic structure of $\rm CuFe_2O_4$ is ferrimagnetic below $\rm
T_c\sim 750K$. If annealed in air and slowly cooled from $\rm 740
^\circ C$, it undergoes a tetragonal distortion (c/a=1.09) below
$T=650K$.

The structural phase transition is driven by the cooperative
Jahn-Teller effect that tends to distort the octahedron to lift
the degeneracy of the doubly degenerate electronic ground-state of
the $\rm Cu^{2+}$ ions. The valence state of copper is +2, so that
every Cu carries an effective spin S=1/2. Fe is in the +3
oxidation state in $\rm CuFe_2O_4$ and, accordingly, has a spin
S=5/2. To obtain the spectrum of magnetic excitations, spin-wave
theory can be used. The Heisenberg operator must now take into
account that $\rm CuFe_2O_4$ has two magnetic sublattices with
different spins, $S_A\sim 1/2$ and $S_B\sim 5/2$,
\begin{equation}
H=\sum_{i,j}J^A_{i,j}\vec S^A_i\cdot \vec
S^A_j+\sum_{i,j}J^B_{i,j}\vec S^B_i\cdot \vec S^B_j
+\sum_{i,j}J^{AB}_{i,j}\vec S^A_i\cdot \vec
S^B_j+\sum_{i,j}J^{AB}_{i,j}\vec S^B_i\cdot \vec S^A_j
\end{equation}
The equations of motion for the spins in the A- and B-sublattices
are similar to the ones which were derived for the simple
ferromagnet. However, the calculations which lead to the spin-wave
dispersion are rather lengthy and will not be reproduced here. A
complete derivation of the spin-wave spectrum for a two
sublattices ferrimagnet can be found \textit{e.g.} in the book of
Turov \cite{turov}. The result yields
\begin{eqnarray}
\hbar \omega_{1,2}(\vec k)& = & \pm
[(S_B-S_A)J^{A,B}(0)-S_A(J^{A}(0)-J^{A}(\vec k))+S_B(J^B(0)
\nonumber \\ & & -J^B(\vec k))]+{1\over
2}\{[2(S_B+S_A)J^{A,B}(0)-2S_A(J^A(0)-J^A(\vec k)) \nonumber \\ &
& -2S_B(J^B(0)-J^B(\vec k))]^2 -S_AS_B(4J^{A,B}(\vec
k))^2\}^{1/2}. \label{ferridisp}
\end{eqnarray}

\begin{figure}
\begin{center}
\includegraphics*[scale=0.35]{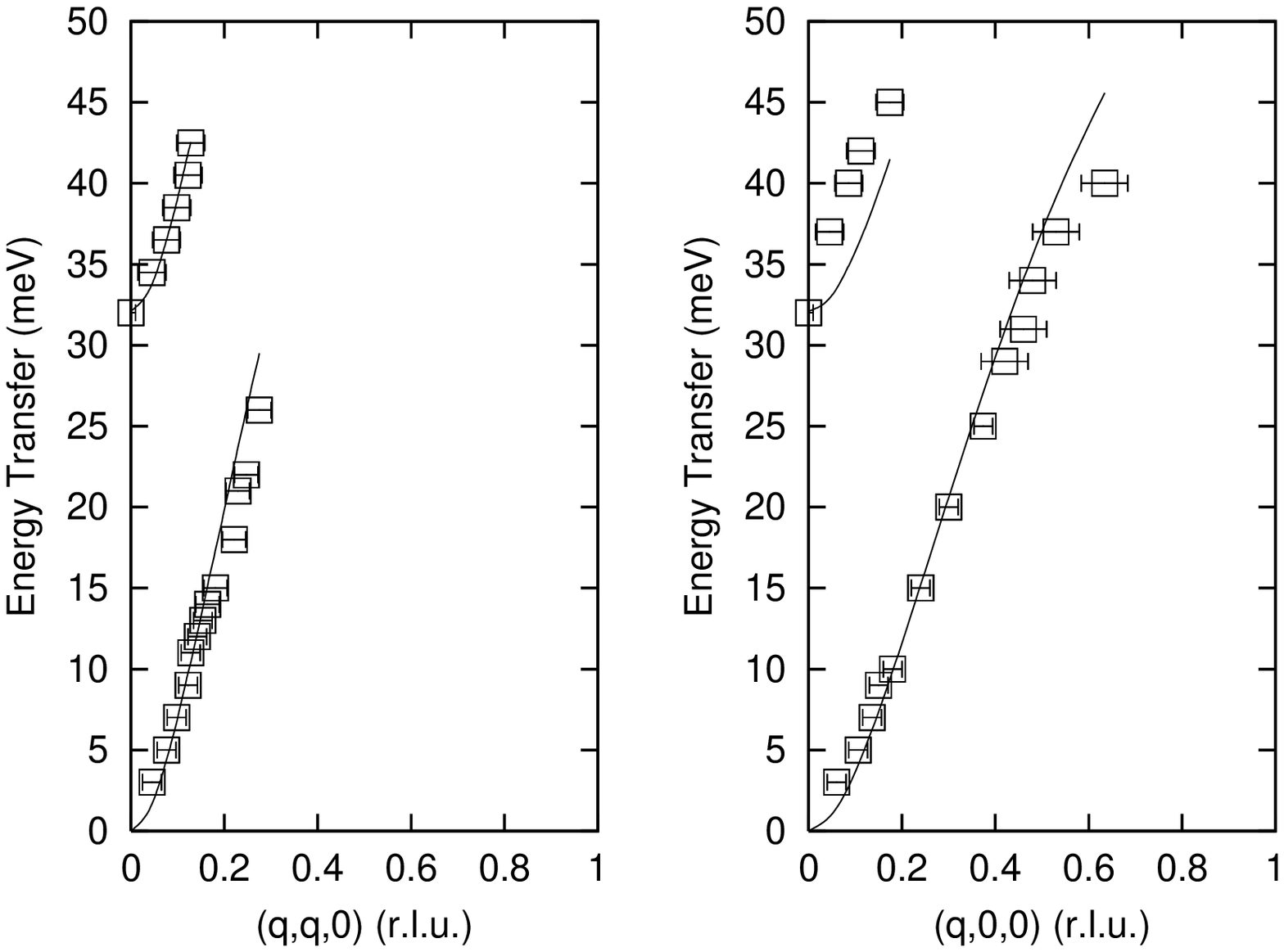}
\kern 0.5cm
\includegraphics*[scale=0.35]{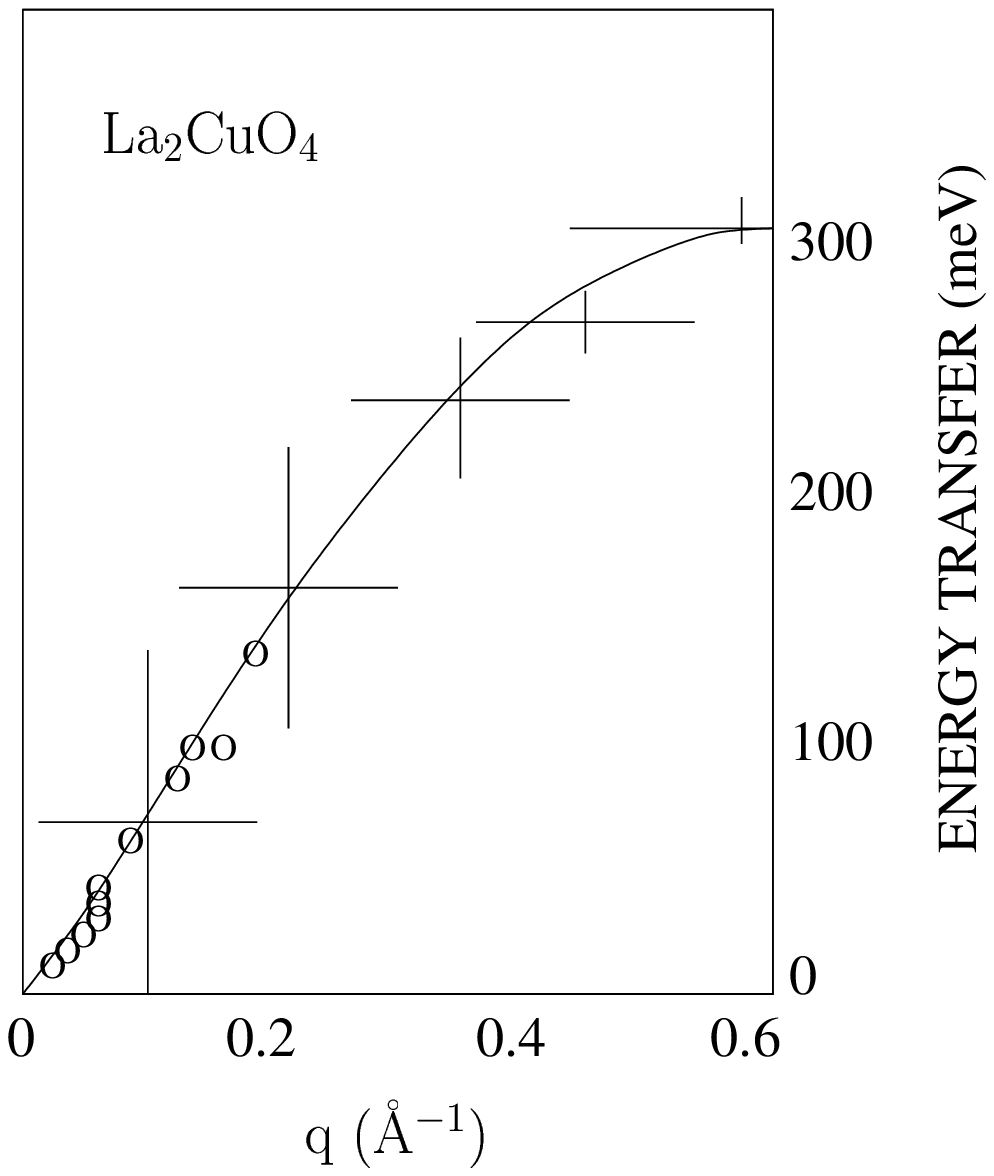}
\vskip 4pt \caption{\textit{Left, Center}: Dispersion of the spin
waves in $\rm CuFe_2O_4$ at T=300K. The line is the result of
calculations using Eq. \ref{ferridisp} with $\rm J^{A,B}=-1.2 meV$
and $\rm J^{A}=J^{B}\sim 0 meV$. (Ref.~ \cite{roessli}).
\textit{Right}: Dispersion of the spin waves in $\rm La_2CuO_4$
(Taken from Ref.~\cite {hayden}).} \label{cufe2o4}
\end{center}
\end{figure}

Two branches of spin waves appear in the spectrum of excitations
in a ferrimagnet. At the magnetic zone center, one branch exhibits
a gap, while the other one goes to zero like $k^2$. Setting
$S^A=S^B=S$ and $J^A=J^B=J$ in Eq. \ref{ferridisp}, the dispersion
relation for a simple collinear antiferromagnet with next-nearest
neighbor exchange interactions is obtained,
\begin{equation}
\hbar \omega_{1,2}(\vec k)=2S\sqrt{(J^{A,B}(0)-J(0)-J(\vec
k))^2-[J^{A,B}(\vec k)]^2} \label{antidisp}
\end{equation}
From Eq. \ref{antidisp} it is seen that the spectrum of the
magnetic excitations of an isotropic antiferromagnet is doubly
degenerate and goes to zero at the magnetic zone center. $\rm
La_2CuO_4$ is a  typical example of an Heisenberg antiferromagnet.
The dispersion of the magnetic excitations in this compound is in
agreement with Eq.~\ref{antidisp}, as shown in Fig.~\ref{cufe2o4}.

\subsection{The role of the anisotropy}

\begin{figure}
\centering
\includegraphics*[scale=0.5]{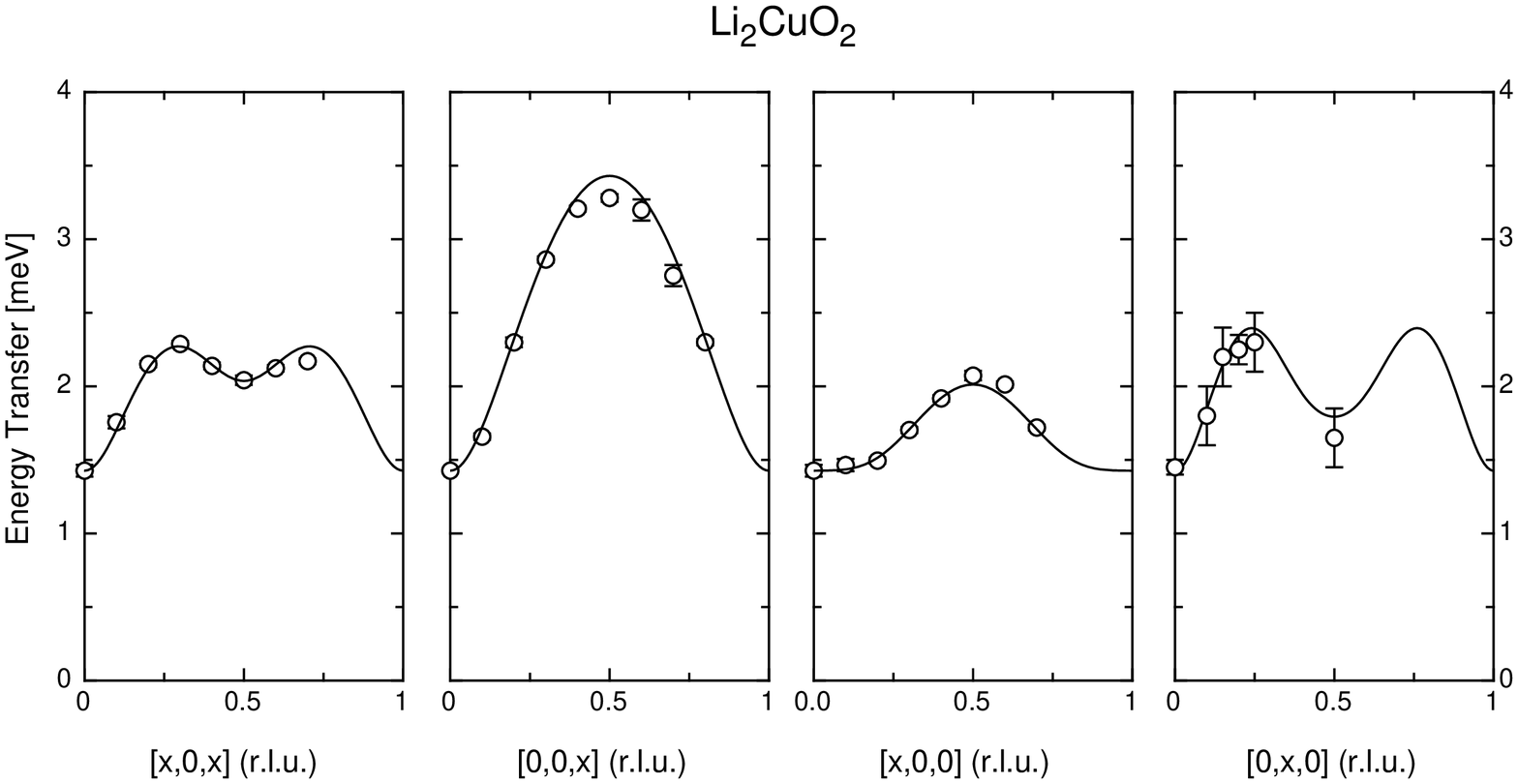}
\vskip 4pt \caption{Dispersion of the spin waves in $\rm
Li_2CuO_2$ at T=1.5K (After Ref. \cite{boehm}).} \label{li2cuo2}
\end{figure}

The spin-spin interactions between electrons like the direct
exchange in ferromagnets and the 'super'-exchange interactions in
antiferromagnets are described by an effective bilinear Hamilton
operator  where the exchange integrals are only functions of the
distance between the ions, $J(\vec r_i-\vec r_j)=J(|\vec r_i-\vec
r_j|)$. However, higher-order terms in the spin Hamiltonian are
possible \cite{tyablikov} and, in general, the spin-spin
Hamiltonian can be expanded in a power series of the spin
operators
\begin{equation}
H=\sum_{i,j}J^{\alpha,\beta}_{i,j}S^{\alpha}_i S^{\beta}_j +
\sum_{i,j,k,l}J^{\alpha,\beta, \delta, \gamma}S^{\alpha}_i
S^{\beta}_j S^{\delta}_k S^{\gamma}_l + \dots
(\alpha,\beta,\delta, \gamma=x,y,z) \label{power}
\end{equation}
Which terms are of significance for a particular physical problem
depends on the lattice symmetry and most often higher order powers
in Eq. \ref{power} give only a small contribution to the magnetic
energy. We show in Fig. \ref{li2cuo2} the result of measurements
of the spin-wave excitations in $\rm Li_2CuO_2$ which is a
three-dimensional collinear antiferromagnet below $\rm T_N=9.2K$.
The model Hamiltonian used to describe the dispersion of the spin
waves in $\rm Li_2CuO_2$ consists of the Heisenberg operator
including uniaxial anisotropy along the spin direction
\begin{equation}
H=\sum_{i,j}J_{i,j} \vec {S}_i  \cdot \vec {S}_j +
\sum_{i,j}J^{\bot}_{i,j}S^{z}_i S^{z}_j.
\end{equation}
The corresponding dispersion relation of the spin waves is given
by $\hbar \omega (\vec k)=S\sqrt{(I(\vec
k)+J(0)-I(0)+D)^2-J^2(\vec k)}$ with $D=J^{\bot}(0) - I^{\bot}(0)$
where $J(\vec k)$ and $I(\vec k)$ are the inter- and
intra-sublattice exchange integrals, respectively~\cite{boehm}. A
single spin-wave branch characterized by a large energy-gap is
obtained. The gap in the spectrum of the magnetic excitations of
$\rm Li_2CuO_2$ is the result of the presence of the strong
anisotropy which aligns the magnetic moments along the easy-axis
of magnetization. It has  been shown that magnetic anisotropy in
this case results from second order spin-orbit coupling which
induces anisotropic super-exchange interactions  in $90^\circ$
Cu-O-Cu bonds as it is realized in $\rm Li_2CuO_2$ and in cuprates
with edge-sharing $\rm CuO_2$ chains \cite{yushankhai}.

Other sources of magnetic anisotropies  originate from the
electric field produced by the atoms surrounding a  magnetic ion.
The form of the crystal field Hamiltonian depends on the symmetry
of the lattice. For tetragonal symmetry, the single-ion anisotropy
has the form $H_A=D\sum_{j}[(S_j^z)^2-{1\over 3}S(S+1)]$. Moriya
has shown that the spin-orbit coupling introduces an effective
interaction between the spins of the form $H_{DM}=\vec D \cdot
\sum_{i,j}(\vec S_i \times \vec S_j)$ in compounds without
inversion center. In some special cases, the anisotropy might be
so large, that the magnetic properties of a compound are well
described by either an Ising-like ($H=\sum_{i,j}S^z_iS^z_j$) or an
XY-like ($H=\sum_{i,j}(S^x_iS^x_j+S^y_iS^y_j)$) model. Table
\ref{table1} gives a summary of typical substances and of the
corresponding models used to account for their magnetic
properties.

\begin{table}[htb]
\caption{Some compounds that show characteristic Heisenberg, Ising
or XY-behavior (after Reference \cite{jaeger}). D is the dimension
of the lattice (D=3: three-dimensional lattice; D=2: layered
structures; D=1: spin chains)} \label{table1}
\begin{center}
\begin{tabular}{l|c|c|c}
\hline
    D   &Heisenberg             &Ising                          &XY   \\
\hline \hline
    3   &EuO,EuS,           &$\rm Tb(OH)_3$,$\rm DyPO_4$,         &$\rm Co(C_5H_5NO_6)(ClO_4)_2$ \\
    {}  &$\rm RbMnF_3$,$\rm KMnF_3$ &$\rm FeF_2$                  &$\rm Fe[Se_2CN(C_2H_5)_2]_2Cl$       \\
\hline
    2   &$\rm K_2CuF_4$,$\rm BaMnF_4$   &$\rm FeCl_2$,$\rm CoCs_3Br_5$,       &$\rm CoBr_2 \cdot 6H_2O$ \\
    {}  &$\rm K_2NiF_4$         &$\rm RbCoF_4$                & $\rm CoCl_2$    \\
\hline
        1       &$\rm (C_6H_{11}NH_3)CuBr_3$      &$\rm CoCl_2 \cdot 2H_2O$                  &$\rm(CH_3)_4NNiBr_3$ \\
        {}      &$\rm RbNiCl_3$,$\rm CsNiCl_3$  &$\rm CsCoCl_3$,$\rm RbFeCl_5 \cdot 2H_2O$ &$\rm Cs_2CoCl_4$,$\rm PrCl_3$ \\
\hline
\end{tabular}
\end{center}
\end{table}

\section{Excitations in itinerant electron magnetism}

The Heisenberg model is based on the assumption that the magnetic
moments are localized around the ions and that the value of the
spin is a multiple half-integer of a Bohr magneton $\mu_B$.
However, metallic compounds like Fe, Co or Ni have magnetic
moments of $\sim 2.12 \mu_B$, $\sim 1.57 \mu_B$ and $\sim 0.6
\mu_B$, respectively. To account for this effect, it must be
recognized that in a metal electrons are 'delocalised' and arrange
themselves in the lattice in electronic bands. This corresponds to
the case  where in the Hubbard model the kinetic energy $t$ is
much larger than the inter-site Coulomb energy $U$. Although the
$(t,U)$ phase-diagram in three-dimensions is not exactly known, it
can be shown that in the limit where $t \gg U$ and within the
mean-field approximation, the Hubbard model reproduces the Stoner
theory of ferromagnetism~\cite{fazekas}.

\subsection{Stoner Theory}

The Stoner theory proceeds in a similar way as the Pauli theory of
paramagnetism. However, it includes the Coulomb repulsion. The
energy of an electron in the presence of an external magnetic
field $H$ and of exchange interactions $U_{ex}$ is given by
$\epsilon_k={{\hbar^2 k^2}\over{2m}} \mp {1 \over
2}(U_{ex}M+gH\mu_B)$, where $M=n_\uparrow-n_\downarrow$ is the
magnetization and $\mp$ refers to the band with spin-up and -down,
respectively. The energy bands are split by an energy $U_{ex}M
+H\mu_B$, which in the absence of a magnetic field is proportional
to the magnetization. This is called the exchange splitting. As a
response to the action of both the external field and of the
exchange interaction, the system develops a spin polarization. The
static susceptibility $\chi = {{\partial M}\over{\partial H}}$ is
accordingly given by \cite{fazekas}:
\begin{equation}
\chi = {{(g \mu_B)^2 \rho (\epsilon_F)}\over {2}} \cdot {{1}\over
{1-U_{ex}\rho(\epsilon_F)}} = {{\chi_{Pauli}}\over{1-U_{ex}\rho
(\epsilon_F)}}. \label{statsusc}
\end{equation}
$\rho(\epsilon_F)$ is the density of the electrons at the Fermi
level. $S_{Stoner}=1/(1-U_{ex}\rho(\epsilon_F))$ is called the
\textit{'Stoner enhancement factor'}. The susceptibility of an
interacting electron gas is enhanced by the factor $S_{Stoner}$
compared to the Pauli susceptibility $\chi_{Pauli}$ of the free
electron system. When $U\rho(\epsilon_F)\approx 1$ the
susceptibility diverges and a phase transition to an ordered
ferromagnetic state occurs. A calculation based on electronic band
calculations taking into account exchange correlation energies is
shown in Table \ref{table2}. The results show that ferromagnetism
in Fe, Co and Ni is explained within the frame of the Stoner
theory.  A great success of the Stoner theory is that it relates
both the value of the magnetic moment at $T=0K$ and the Curie
temperature $T_c$ to the enhancement factor $S_{Stoner}$,
\begin{equation}
M(0)=\sqrt{-S_{Stoner} \rho^3(\epsilon_F)/a}
\end{equation}
\begin{equation}
T_c={\sqrt{3}\over{\pi k_B}}\sqrt{-S_{Stoner}\rho(\epsilon_F)/a},
\end{equation}
with $a$ the lattice constant. Accordingly, materials with a large
Stoner enhancement factor are expected to order ferromagnetically
at high temperatures and to have a large magnetization at
saturation, in agreement with experimental values tabulated in
Table \ref{enhancement} for some typical metallic compounds.

\begin{table}
\caption{Stoner enhancement factor from electronic
band-calculations (after Refs. \cite{jaeger},\cite{harrison})}
\label{enhancement} \label{table2}
\begin{center}
\begin{tabular}{p{3cm}|p{2cm}|p{2cm}|p{2cm}}
\hline
     { }           &$U_{ex} \rho(\epsilon_F)$                    &$S_{Stoner}$                  &$T_c$ (K)\\
\hline \hline
     Al           &0.25                                   &1.71                   &-   \\
     $\alpha-$Fe           &1.43                                   &-2.34         &1044 \\
     Co           &1.70                                   &-1.43                  &1390 \\
     Ni           &2.04                                   &-0.97                  &624 \\
     Cu           &0.11                                   &1.12                   &-  \\
     Pd           &0.78                                   &4.44                   & - \\
\hline
\end{tabular}
\end{center}
\end{table}

\subsection{Dynamical susceptibility}

\begin{figure}

\centering
\includegraphics*[scale=0.5]{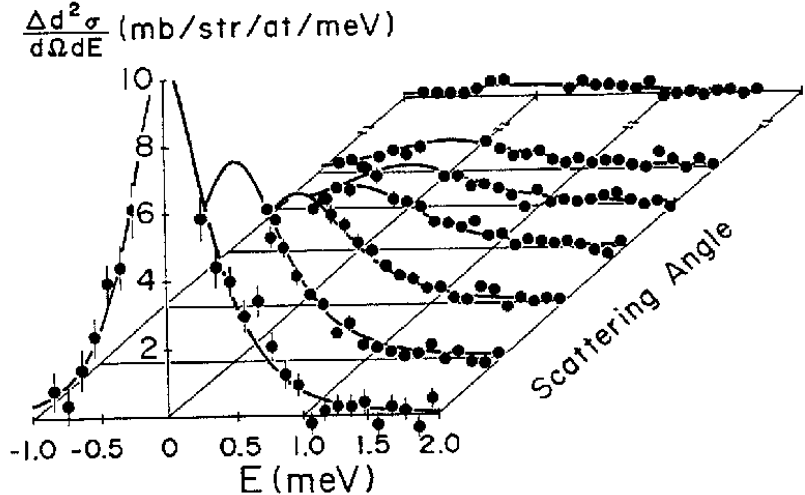}
\vskip 4pt \caption{Energy dependence of the scattering intensity
in the paramagnetic metal $\rm Ni_3Ga$ (taken from Ref.
\cite{bernhoeft}).} \label{ni3ga}
\end{figure}

Materials for which  $U_{ex} \rho(\epsilon_F)$ is close to the
critical value of 1 (like Pd) are called \textit{'nearly
ferromagnets'}.  The imaginary part of the dynamical
susceptibility is enhanced by the exchange interaction, so that
fluctuations in such paramagnetic metals are directly accessible
by inelastic neutron scattering as,
\begin{equation}
 \Im \chi(\vec k, \omega)=
\omega \chi(\vec k){{\Gamma (\vec k)}\over{\omega^2 +
\Gamma^2(\vec k)}}, \label{enhancedsusc}
\end{equation}
where $\chi^{-1} (\vec k)= \chi^{-1}+ck^2$ is the inverse static
susceptibility and $\Gamma(\vec k)=\gamma\chi(\vec k)$ is a
characteristic relaxation frequency. The microscopic constants $c$
and $\gamma$ characterize the static and dynamic parts of the
dynamical susceptibility and can be calculated from the
electronic-band structure \cite{lonzarich}. $\rm Ni_3Ga$ is a
classical example of a paramagnetic metal with strongly correlated
3d transition electrons and exchange enhanced susceptibility. The
imaginary part of the dynamical susceptibility as measured with
inelastic neutron scattering is consistent with Eq.
\ref{enhancedsusc}, which is characterized by a strong rise of the
scattering intensity at low $k$ and $\omega$ \cite{bernhoeft}, as
shown in Fig. \ref{ni3ga}.

\begin{figure}
\begin{center}
\includegraphics*[scale=0.55]{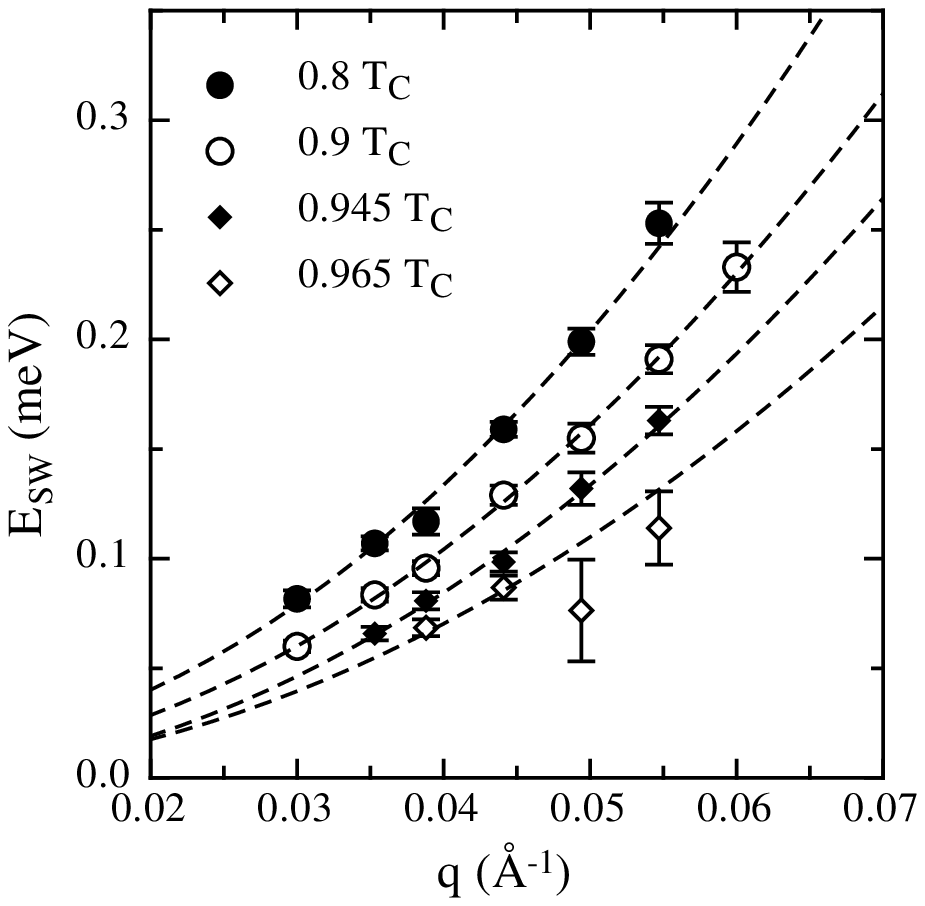}
\kern 0.1 cm
\includegraphics*[scale=0.55]{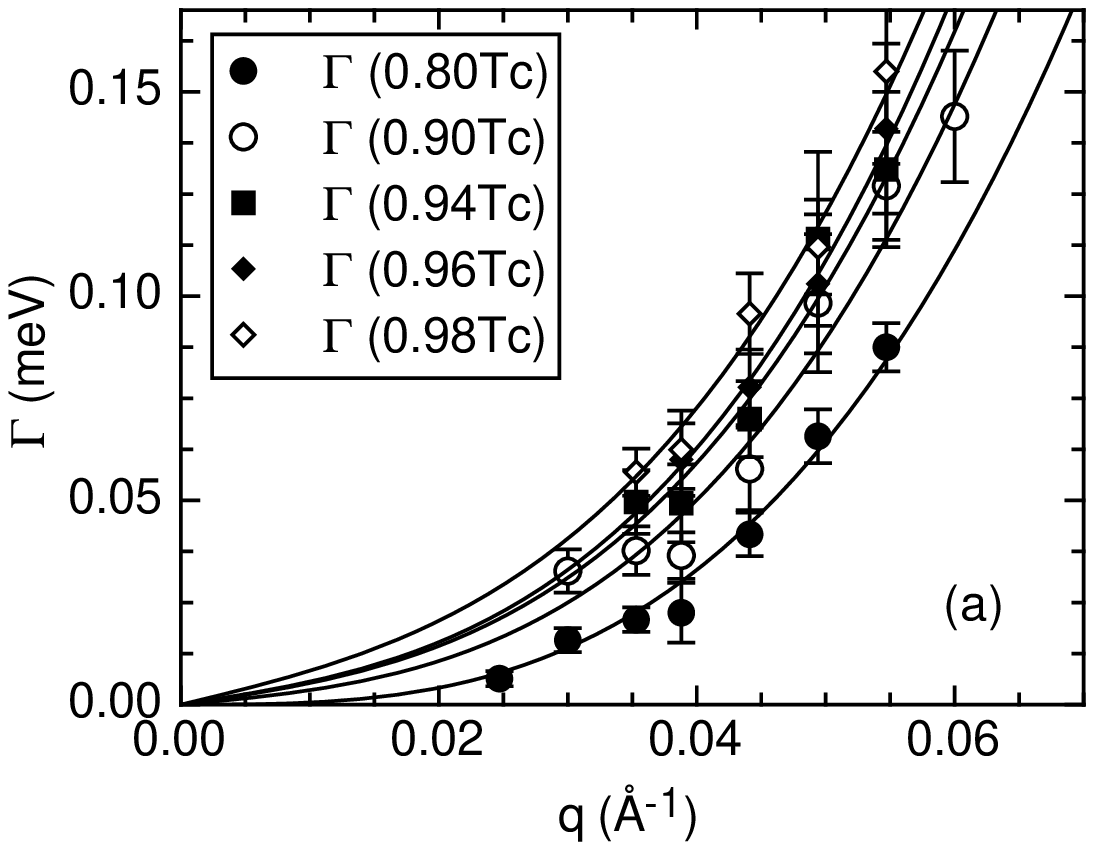}
\vskip 4pt \caption{\textit{Left}: Dispersion of the magnetic
excitations in $\rm Ni_3Al$ as a function of temperature.
\textit{Right}: Dependence of the line-width of the spin-wave
excitations in $\rm Ni_3Al$ as a function of momentum transfer.
See Ref.~\cite{semadeni} for details.} \label{ni3al}
\end{center}
\end{figure}

$\rm ZnZr_2$ is characterized by an exchange enhancement factor
that is just above the critical value, namely
$U_{ex}\rho(\epsilon_F)=1.015$. $\rm ZrZn_2$ orders
ferromagnetically at $\rm Tc\sim 25K$ and the magnetic moment at
saturation is $\mu=0.12\mu_B$. There exists a large class of such
materials which order at low temperature with very small ordered
magnetic moments called \textit{'Weak Ferromagnets'}. For these
materials, and in the ferromagnetic phase, the spectrum of
magnetic excitations consists of well-defined spin waves at small
values of $\vec k$ that merge into a continuum of excitations
(\textit{'Stoner Continuum'}) at large values of momentum
transfers. The dispersion of the spin-waves is quadratic $\hbar
\omega(\vec k)=Dk^2$. D is the stiffness constant and is
proportional to the magnetization, namely
$D=2\mu_BcM$~\cite{moriya}. As an example, Fig. \ref{ni3al}, shows
the dispersion of the magnetic excitations in $\rm Ni_3Al$ that is
also a typical weak ferromagnet with $T_c=72K$ and $\mu=0.075
\mu_B$.\\

\begin{figure}
\begin{center}
\includegraphics*[scale=0.325]{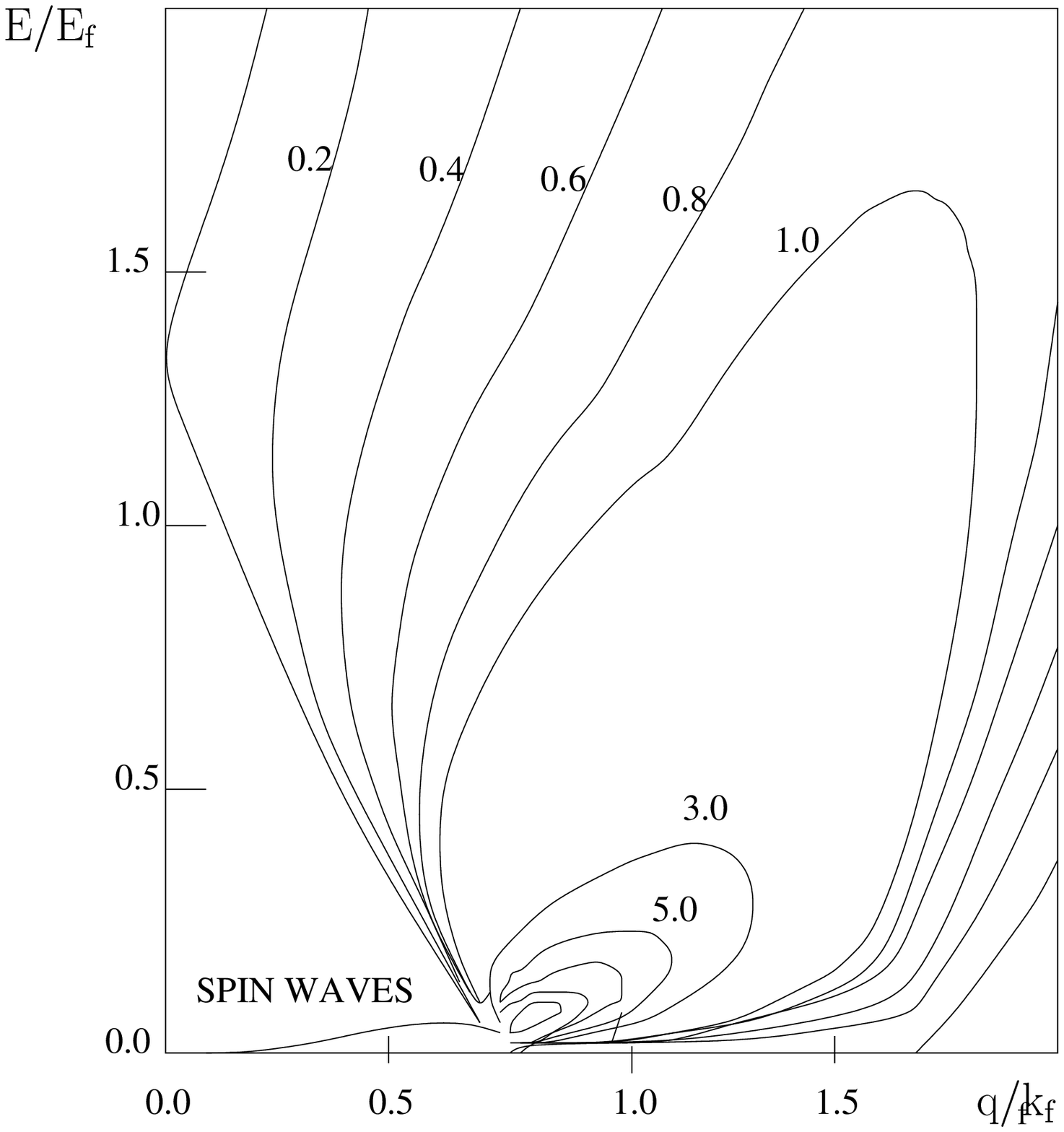}
\kern 0.5cm
\includegraphics*[scale=0.35]{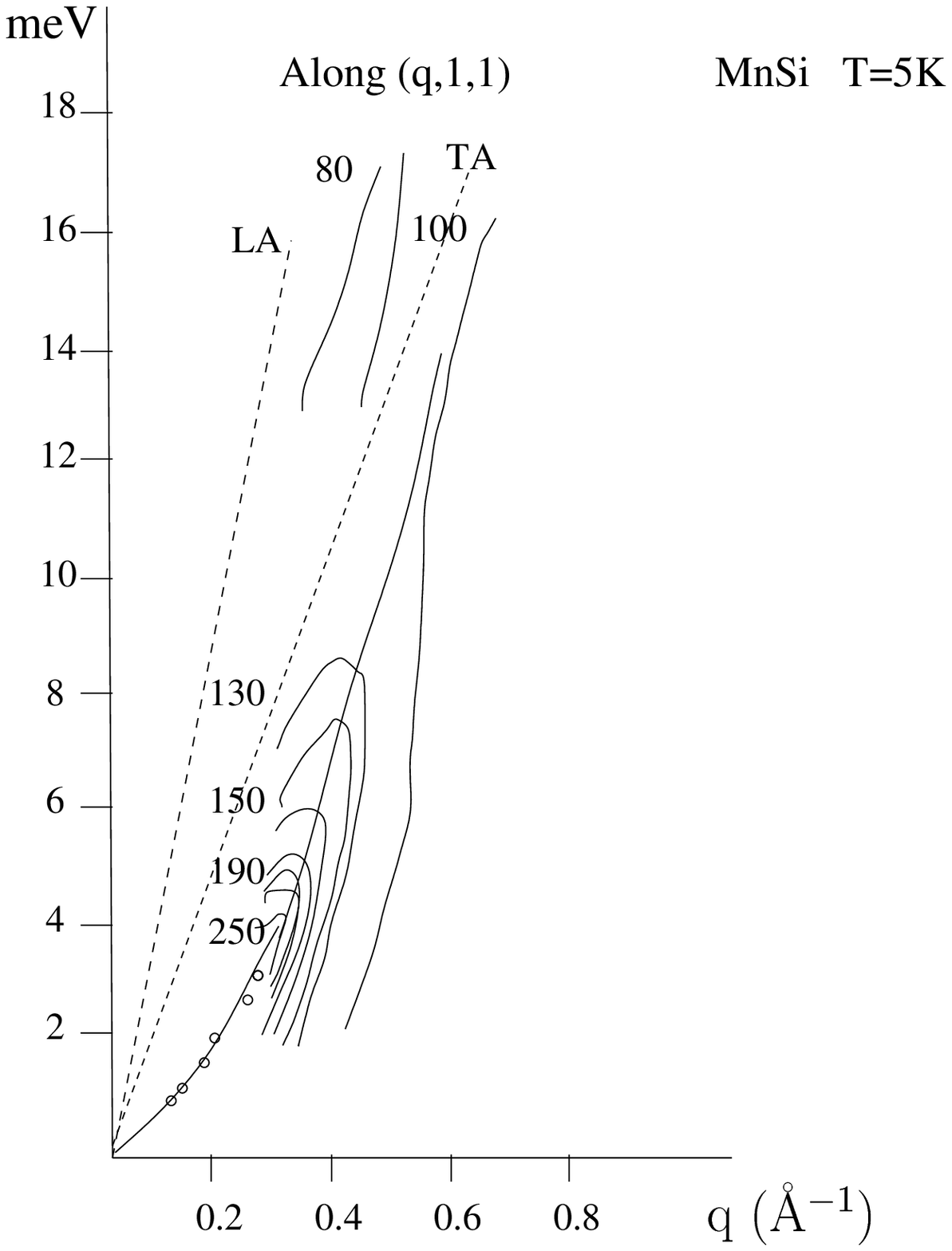}
\vskip 4pt \caption{\textit{Left}: Dispersion of the magnetic
excitations in a ferromagnetic gas. The contour lines refer to the
imaginary part of the dynamical susceptibility. \textit{Right}:
Dispersion of the magnetic excitations in MnSi at T=5K and
H=10kOe. Reproduced from Ref. \cite{ishikawa}. Note the quadratic
dispersion of the spin waves and the Stoner continuum above $\rm
\hbar \omega \sim 3meV$.} \label{stoner}
\end{center}
\end{figure}

The neutron cross-sections for magnetic excitations in the Stoner
continuum are quite involved and can be found in
Ref.~\cite{lovesey}. Fig. \ref{stoner} shows the dispersion and
the intensity of a ferromagnetic electron gas. It is seen, that
the neutron intensity decreases upon entering the Stoner continuum
and the line shape of the magnetic excitations significantly
increases. This is in agreement with neutron scattering studies in
Fe, Ni \cite{mook} and MnSi \cite{ishikawa}.

\subsection{Generalized Susceptibility}

The Stoner theory is based on the free electron model with Coulomb
interactions. For real shapes of the Fermi surfaces in metals, the
static susceptibility of Eq. \ref{statsusc} is $\vec k$-dependent
\begin{equation}
\chi(\vec k)=(g\mu_B)^2{\chi^0(\vec k) \over {1-U_{ex}\chi^0(\vec
k)}},
\end{equation}
where $\chi^0(\vec k)$ is not the Pauli-susceptibility anymore but
depends on the details of the electronic band structure
\cite{overhauser}. Similarly to the case of ferromagnetism, the
system can undergo a phase transition if $U_{ex}$ is large enough.
If the Lindhard-function for free electrons in three-dimensions is
chosen for $\chi^0(\vec k)$, the generalized static susceptibility
diverges for $\vec k \rightarrow 0$ that corresponds to the on-set
of ferromagnetism. In the general case, the susceptibility can
diverge at any value of $\vec k$, which leads to the formation of
spin-density waves. A well-known example is chromium which
undergoes a phase-transition to an incommensurate phase at $\rm
T_N=311K$. The enhancement factor $1-U_{ex}\chi^0(\vec k) \approx
1.025$ is only slightly above the critical value.  The
incommensurate spin-density wave structure of Cr associated with
wave vector $\vec k_0=(0,0,\sim 0.96)$ can be ascribed to nesting
of the Fermi surface~\cite{lomer}. The theory of the magnetic
excitations in weakly antiferromagnetic and helimagnetic metals
was developed by Moriya and is similar to the ferromagnetic case.
We refer to the book of Moriya for details~\cite{moriya}.

\section{Conclusions}

\label{conclusions}

Inelastic neutron scattering is a powerful means for the
investigation of spin waves and spin fluctuations in magnetic
materials. Starting with the Heisenberg model we have shown that a
measurement of the dispersion relation of the spin waves allows a
direct determination of the exchange constants between the
magnetic moments. Close to the critical temperature the spin
deviations from the magnetisation direction $\vec M$ become large
leading to a damping of the spin waves and thus to the appearance
of longitudinal fluctuations along $\vec M$. These modes can be
separated by means of polarisation analysis.

In general, crystal field effects, dipolar and spin-orbit
interactions etc. lead to higher order terms in the Hamiltonian
which can lead to the appearance of an energy gap and/or lift the
degeneracy of the spin-wave modes. Systems with more than one
magnetic moment per unit cell show several dispersion branches.

The picture of localised moments breaks down as soon as the
carriers of the magnetic moments interact or become part of the
sea of conduction electrons. Here, a description of the spin
fluctuations in terms of itinerant models is more appropriate.
Neutron scattering shows directly that spin waves exist at small
$\vec k$ merging into a sea of single-particle (Stoner)
excitations at larger $\vec k$. In nearly ferromagnetic materials
the low energy fluctuations are dramatically enhanced.

In summary, inelastic neutron scattering is at present the only
method to measure spin fluctuations over a large range of $\vec k$
and $\omega$ in localised and itinerant systems. It has provided
invaluable input for a better understanding of the interactions in
weakly and strongly correlated materials that are of current
interest in the field of high-$T_c$ superconductiors, colossal
magneto-resistance materials and heavy-fermion system. Due to
limitations in space we refer the reader to the specialized
literature on these subjects.

\end{document}